\documentclass[11pt]{iopart}
\usepackage{color}
\usepackage{cite}
\usepackage[colorlinks=true,linkcolor=blue,urlcolor=blue,citecolor=blue]{hyperref}
\usepackage{upgreek}
\usepackage{footmisc}

\usepackage[colorinlistoftodos]{todonotes}

\begin{document}

\title[]{Design of a Miniature Kibble Balance for Kilogram-Scale Mass Calibration -- KBmini}

\author{Shisong Li$^\dagger$, Nanjia Li, Weibo Liu, Elsayed E.E. Qupasie,\\ Wei Zhao, Songling Huang}

\address{Department of Electrical Engineering, Tsinghua University, Beijing 100084, China}
\ead{shisongli@tsinghua.edu.cn}
\vspace{10pt}
\begin{indented}
\item[]Present version 01/2026.
\end{indented}

\begin{abstract}
Tabletop version Kibble balances are a significant developing trend for mass realizations following the revised International System of Units. A key innovation through the miniaturization of the Kibble balance from a large-scale instrument into a tabletop device is making the quantum-based realization of mass accessible to a wider range of calibration laboratories and industries. This paper presents a tabletop Kibble balance design at Tsinghua University targeting E2-accuracy class mass calibrations from {1\,g to 1\,kg. For calibrating a mass of 1\,kg, for instance, the required relative standard measurement uncertainty must be below 0.27\,ppm to meet E2-accuracy class.} Major components and features of the proposed system are discussed. A novel {method of multi-harmonic excitation} is proposed to improve the coil-motion linearity during velocity measurement. {We show} that injecting odd-order harmonics into the motion-driving current can significantly improve the uniformity of the coil's moving velocity, while the second-order component can address the asymmetry between upward and downward movements. {This achieves a flat velocity $\Delta v/v< 5\%$ over 60\% of the motion cycle.}
\end{abstract}



\submitto{\MET}


\clearpage
\begin{flushleft}

\section{Introduction}
With the adoption of the revised International System of Units (SI) at the 26th General Conference on Weights and Measures (CGPM) in 2018~\cite{CGPM2018}, fundamental metrology has entered a new era where base units are defined by invariant physical constants. In the new SI, the unit of mass, the kilogram, is defined based on the Planck constant \( h \), where $h$ = 6.62607015 $\times 10^{-34}$\,{{J\,s}} is fixed exactly. Two primary experimental methods link macroscopic mass measurements to \( h \), i.e., the X-ray Crystal Density (XRCD) method (also known as the International Avogadro Cooperation -IAC project)~\cite{XRCD2017}, and the Kibble balance~\cite{Kibble1976,Robinson_2016watt}.
The Kibble balance, in particular, has become a major focus of research at many mass metrology laboratories worldwide as mass realization instruments following the revised SI~\cite{NRC2017,NIST2017,METAS2022,LNE2025,Fang_2020BIPM,NIM2023,MSL2020,KRISS2020,UME2023,NPLtabletop2024,PTB_PB2_2021,THUdesign2022}.

The Kibble balance is an instrument that compares the mechanical power to electrical power, which was first proposed by {Dr.} Bryan Kibble at the National Physical Laboratory (NPL, UK) in 1976~\cite{Kibble1976}. The operation of a Kibble balance experiment relies on two distinct phases: the weighing phase and the velocity phase. The weighing phase compares the Lorentz force generated from a current-carrying coil immersed in a magnetic field to the gravitational force of a test mass, written as $BlI=mg$, where $B$ is the magnetic flux density at the coil position, $l$ the coil wire length, $I$ the DC current passing through the coil, $m$ the mass to be measured, and $g$ the local gravitational acceleration. 
While the velocity phase measurement is operated under governance of Faraday’s law for coil induction. The same coil is moved vertically at a speed $v$ in the same magnetic field, and an induced voltage $U=Blv$ is obtained. By eliminating the same geometrical factor, $Bl$, the test mass value is determined as $m=UI/(gv)$.
To realize $m$ in terms of the Planck constant $h$ and achieve a {relative measurement uncertainty} in the order of $10^{-8}$, quantum standards, such as the quantum Hall resistance standard and the Josephson voltage standard should be employed, as detailed in \cite{Haddad_Bridging2016}, {otherwise, an uncertainty on the order of $10^{-6}$ is achievable.}

In light of the substantial size and weight of the experimental setup of precision Kibble balances, e.g.\cite{NIST2017,NRC2017}, which are typically large and complicated, a growing research focus has shifted towards developing compact-sized tabletop Kibble balances. Initiatives are underway at institutions including the National Institute of Standards and Technology (NIST, USA)~\cite{NIST_QEMMS,NISTtabletop2020,NISTtabletop2019IEEE,NISTTabletop2024}, the NPL~\cite{NPLtabletop2024,NPL2022}, and the Physikalisch-Technische Bundesanstalt (PTB, Germany) in collaboration with Ilmenau University of Technology~\cite{PTB2018,PTB_PB2_2021,PTB2025}. In late 2022, Tsinghua University also launched a tabletop Kibble balance project targeting NMI (National Metrology Institute)-level, E1-accuracy class mass calibration~\cite{THUdesign2022}, with subsequent progress reported in~\cite{THUKB2023, li2024magnet, THU2024currentsource, THU2024gravity, THU2025, THUmag2025,Liu_2025}.

Building on this trend, and inspired by the recent demonstration that a flexure hinge can facilitate velocity measurement via sub-millimeter-range sinusoidal oscillation~\cite{PTB2025}, this paper presents a novel tabletop Kibble balance — KBmini — designed for industrial E2-accuracy class mass calibration in air operation. {A key innovation of the KBmini design is its use of multi-harmonic excitation to achieve sub-millimeter-range motion during the moving phase.} This approach induces a quasi-linear coil motion, enabling uniform velocity control and thereby simplifying the challenges associated with traditional AC electrical measurements in such systems~\cite{PTB2025,MSL2020}. 

The remainder of this paper is structured as follows: Section \ref{sec02} details the overall design of the KBmini system. Section \ref{sec03} presents some major considerations of the proposed experiment based on primary experimental results, focusing on the compact magnet-coil system and {the multi-harmonic excitation technique for realizing} quasi-constant-velocity motion. Finally, conclusions and future research directions are summarized in Section \ref{sec04}.

\section{Overall Design}
\label{sec02}

\begin{figure}
    \centering
    \includegraphics[width=0.7\linewidth]{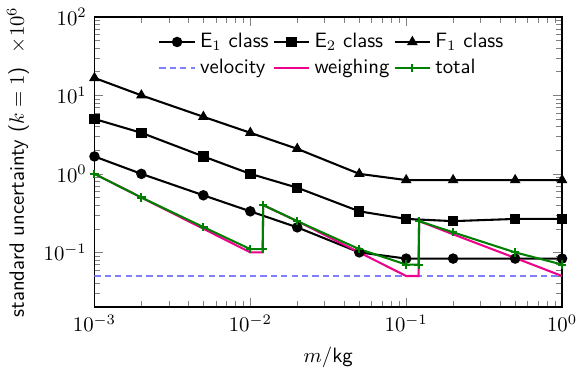}
    \caption{
    The {relative measurement uncertainty} target of KBmini. The black curves show the standard {relative measurement uncertainty} of mass calibrations for E1, E2, and F1 classes. The dashed blue line is the velocity {phase measurement uncertainty}, $5\times10^{-8}$, not depending on the mass measurement range. The magenta curve is the weighing {phase} uncertainty. {The current source has three ranges with matching sampling resistors}, covering measurement ranges of respectively [1\,g,12\,g], [12\,g, 120\,g], and [120\,g, 1000\,g]. {The uncertainty steps are range boundaries, and in each range, the weighing resolution is fixed, and hence the weighing uncertainty increases as the mass value scales down.}
    }
    \label{fig01}
\end{figure}

As shown in figure \ref{fig01}, the design objective of KBmini is to achieve E2-accuracy class mass calibration for a range from 1\,g to 1\,kg. A high $Bl$ product ($\approx 1400$\,Tm, detailed in Section \ref{sec031}) ensures low {relative measurement uncertainty}, on the order of $10^{-8}$. An automatic system for matching the current source range and the current-measuring resistances at 1\,kg, 120\,g, and 12\,g prevents the continuous increase in uncertainty when measuring smaller masses. Furthermore, the design retains E1-accuracy class mass calibration capability at specific points, e.g. 1\,kg, 100\,g, and masses below 10\,g.

\begin{figure}
\centering
\includegraphics[width=0.7\textwidth]{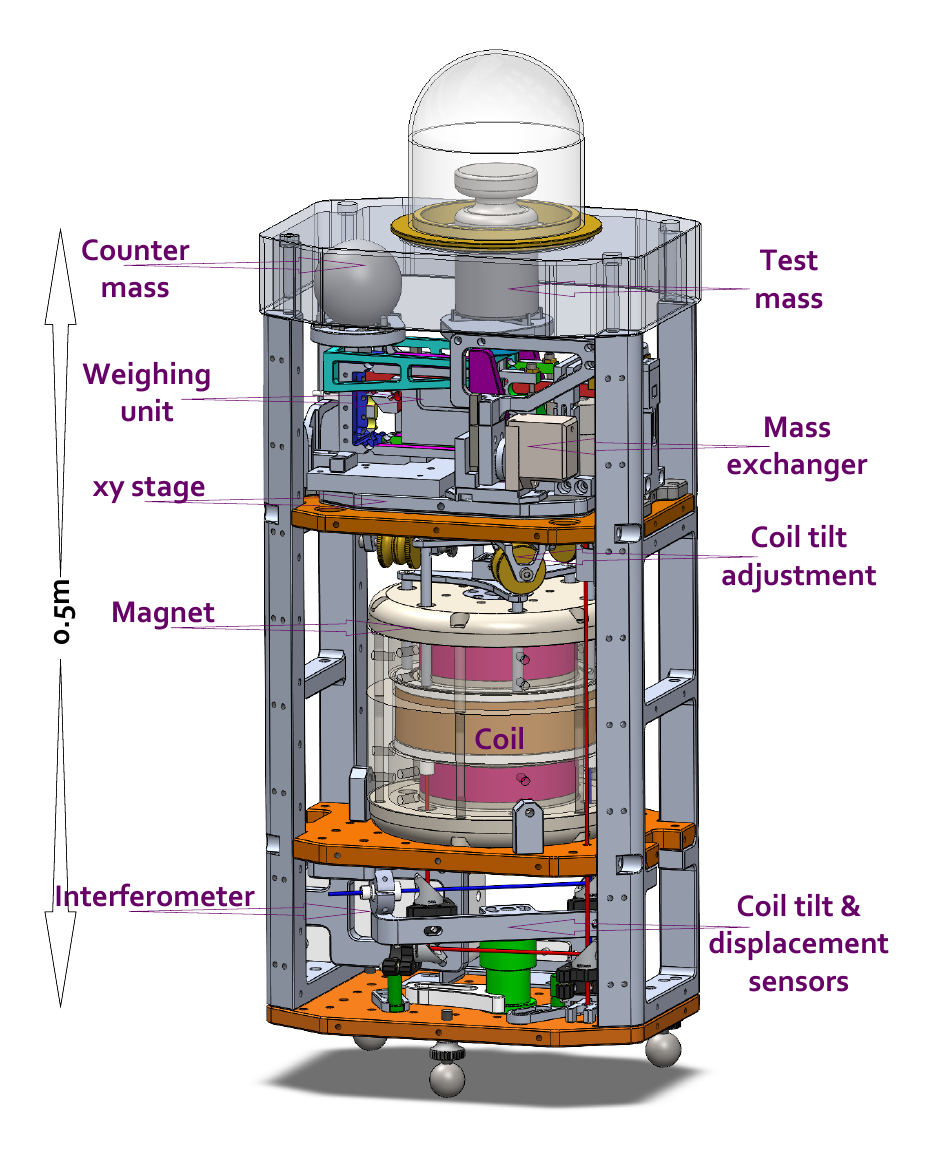}
\caption{The CAD model of KBmini.}
\label{fig02}
\end{figure}

As shown in figure \ref{fig02}, the experimental setup of KBmini includes the following major components: 
\begin{itemize}
    \item \textbf{weighing unit.} A self-developed weighing unit~\cite{THU2025} compares the electromagnetic force to the gravitational force of the mass. Unlike traditional designs that use optical sensors, this unit integrates a capacitive sensor for displacement detection. This approach improves weighing sensitivity, allowing for a stiffer flexure design that can support a higher dead-load capacity. The unit features an independent actuator comprising a BIPM-type magnet and a bifilar coil. The $Bl$ value for each coil is $(Bl)_A\approx30$\,Tm. One coil is short-circuited to provide electromagnetic damping, while the other is used as the drive coil during the moving phase.
    
    \item \textbf{mass exchanger.}  The mass exchange system incorporates two motors mounted next to the weighing unit. One is used to load and unload the test mass $m$, and the other for lifting the $m/2$ counterweight. 
    
    \item \textbf{magnet.} The magnet system is a more compact version of the BIPM openable magnet circuit developed at Tsinghua University~\cite{li2024magnet}. The outer diameter and total height of the magnet are respectively 150\,mm and 135\,mm. The magnetic flux of two NdFeB permanent magnet disks is guided through an air gap of 11\,mm in width, 40\,mm in height and 54.5\,mm in mean radius, achieving a flux density of $\approx 0.6\,$T. The construction of the magnet circuit is detailed in section \ref{sec031}.

    \item \textbf{coil.} The coil implemented in KBmini is characterized by a height of 30\,mm with an average radius of 55.25\,mm. A wire diameter of 0.15\,mm is chosen for the coil winding. The coil contains a total number of turns $N=7000$ with a total resistance of 2.4$\,$k$\Omega$. The magnet and coil yield a $Bl$ value of about 1400\,Tm. 

    \item \textbf{$xy$ adjustment stage.} The $xy$ adjustment stage is mounted directly under the weighing unit and facilitates fine positioning of the coil inside the magnet’s air gap. {The stage comprises a fixed inner frame and a movable outer frame, which carries the whole weighing unit. The inner and outer parts are connected flexibly, and $x$ and $y$ motions can be achieved by adjusting two perpendicular screws.} The adjustable range is $\pm$1\,mm in both the $x$ and $y$ axes.

    \item \textbf{coil tilt adjustment.} The coil tilt adjustment is used to correct the levelness of the coil, ensuring the coil's plane is aligned horizontally. The adjustment is realized by moving the position of three counterweights at every 120 degrees. 

    \item \textbf{interferometer.} The coil displacement and velocity are measured by an {SIOS} SP 5000 DI laser interferometer, which provides a measurement resolution of 5$\,$pm, and an angular range of $\pm$430$\,\upmu$rad with a resolution of 0.005\,$\upmu$rad. The system is powered by a HeNe laser featuring a frequency stability of $2\times10^{-8}$.

    \item \textbf{coil motion sensors.} The parasitic motions ($x$, $y$, $\theta_x$, $\theta_y$) of the coil are measured via an optical sensor system. A flat mirror and a corner-cube retroreflector are mounted beneath the coil to establish the measurement optical path. The laser beams reflected from these two components are captured by two position-sensitive detectors (PSDs). It enables the simultaneous measurement of the coil's tilt angle via the flat mirror beam and its horizontal displacement via the corner-cube retroreflector.

    \item \textbf{current source.} A bipolar current source with high short-term stability was implemented for the KBmini, combining two commercial current sources with different ranges (e.g., $\pm$20\,mA and $\pm$2\,$\upmu$A for 1\,kg mass calibration). The fine source compensates for the variation of the main source, offering enough resolution for the weighing. A comprehensive description of such a system can be found in \cite{THU2024currentsource}. 

    \item \textbf{electrical references.} {Three standard resistors (1\,k$\Omega$, 10\,k$\Omega$, and 100\,k$\Omega$) are used for multi-range current sampling, and a continuously powered Zener reference (e.g., a Fluke 732C), which is regularly calibrated, is employed as the voltage standard.} 

\end{itemize}

As presented in figure \ref{fig03}, the operation of KBmini will employ a typical two-mode two-phase (TMTP) measurement scheme. During the weighing phase, the counterweight $\approx mg/2$ is unloaded, enabling an upward force $F_C\approx mg/2$ on the flexure hinge. During the mass-off measurement, a downward electromagnetic force generated by the coil balances the weighing unit, i.e.
\begin{equation}
    BlI_-+F_C=0,
    \label{massoff}
\end{equation}
where $I_-$ is the balancing current during the mass-off measurement. During the mass-on measurement, the test mass is loaded, and meanwhile, the current in the coil is changed into the opposite direction. The force is balanced as
\begin{equation}
    BlI_++F_C=mg.
    \label{masson}
\end{equation}
A subtraction of two equations, (\ref{massoff}) and (\ref{masson}), yields the Kibble balance weighing equation. A symmetrical current, i.e. $I_+=-I_-=I$, can eliminate the linear current effect~\cite{CoilCurrent_Li2018}.
In the moving phase, the test mass is unloaded, and the counterweight is loaded to rebalance the hinge. Then, a periodic current of 1\,Hz is applied to the weighing unit actuator, allowing it to drive the system into a periodic oscillation. Throughout both phases, the system is maintained in a state of equilibrium, and the sum of all forces remains zero at any given moment.

\begin{figure}
    \centering
    \includegraphics[width=\linewidth]{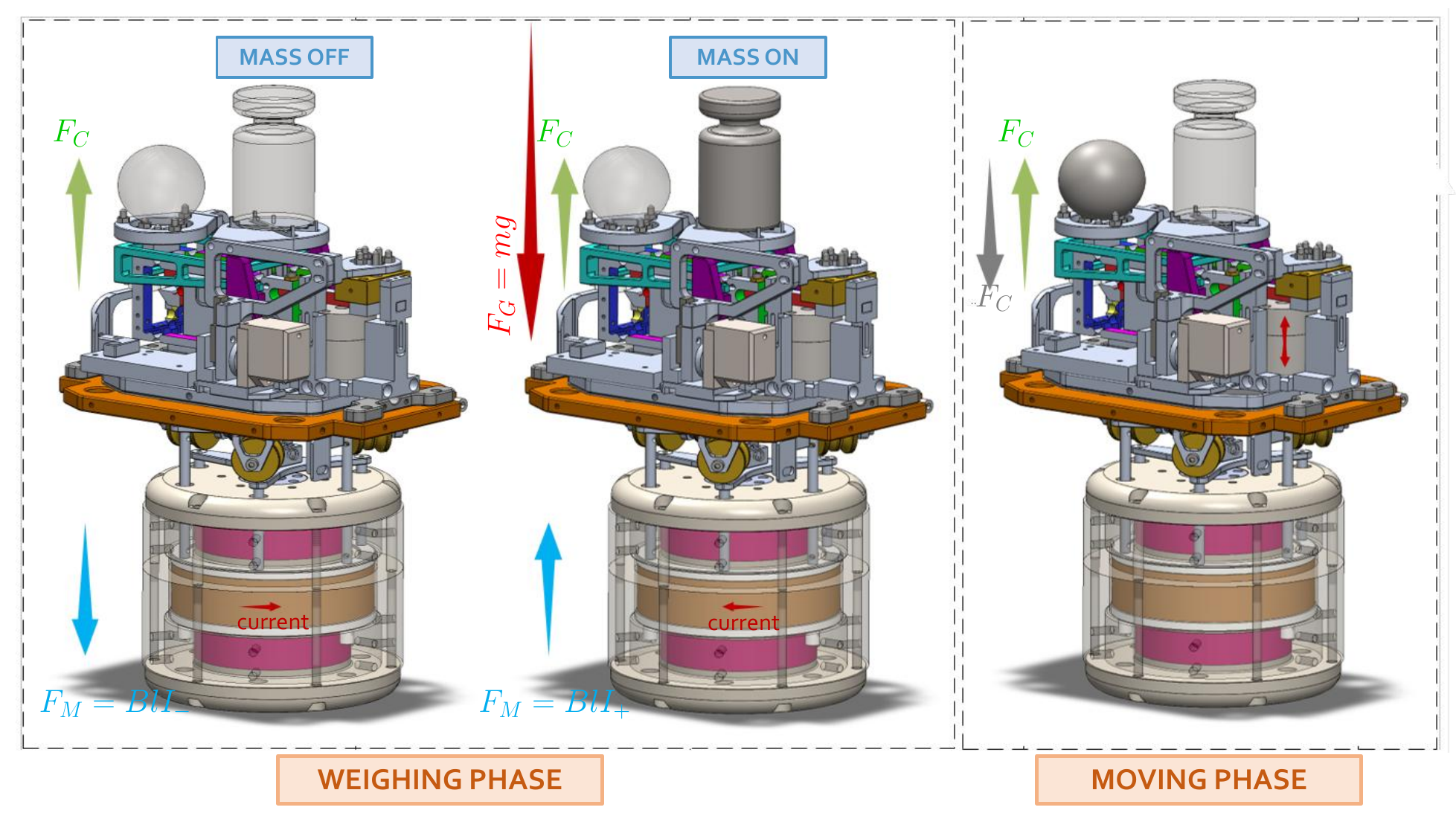}
    \caption{Measurement scheme of KBmini. The weighing measurement consists of two steps: mass-off and mass-on. In the mass-off step, with both the counterweight and test mass unloaded, a downward electromagnetic force \(F_M = BlI_-\) balances the upward unbalanced force from {counterweight \(F_C\)}, establishing equilibrium: {\( -F_C + F_M = 0\)}. During the mass-on step, the test mass is loaded and the coil current is reversed, resulting in a new force balance: {\(-F_C + F_G -F_M = 0\)}, where \(F_G = mg\). Subtracting the mass-off equation from the mass-on equation yields the expression \(Bl = \frac{mg}{I_+ - I_-}\). In the subsequent velocity phase, the counterweight is loaded while the test mass is unloaded; the counterweight is chosen to exert a force \(-F_C\) such that the total static force remains zero. }
    \label{fig03}
\end{figure}

\section{Major features \& considerations}
\label{sec03}

\subsection{Magnet-coil system}
\label{sec031}

The magnetic geometrical factor $Bl$ is implicated in both moving and weighing measurements. Although it is {substituted} and is not present in the final mass determination equation, its {accuracy} is pivotal to the {overall uncertainty of the mass determination}. 

The KBmini system employs a BIPM-type magnet circuit \cite{stock2006watt,Li_2022_irony}, as shown in figure \ref{magnet_ex}(a). This system is a more compact version of the openable magnet circuit developed in \cite{li2024magnet}. The outer diameter is 150\,mm, the total height is 135\,mm, and the total weight is approximately 14\,kg (in contrast, the {magnet system of} Tsinghua tabletop Kibble balance weighs 40\,kg). The upper and lower ring-shape permanent magnet disks (inner radius 10\,mm, outer radius 45\,mm, height 20\,mm) are made of NdFeB, which has a coercivity of $H_c\approx -1000$\,kA/m. This material choice, informed by previous studies, aims to achieve a higher magnetic field strength in the air gap. The yoke material is made of pure iron with high permeability. Their flux of two permanent magnet disks is guided by the yoke through a 11\,mm wide, 40\,mm height, and 54.5\,mm mean radius air gap, generating a flux density of approximately 0.6\,T. 

\begin{figure}[tp!]
\centering
\includegraphics[width=1.05\textwidth]{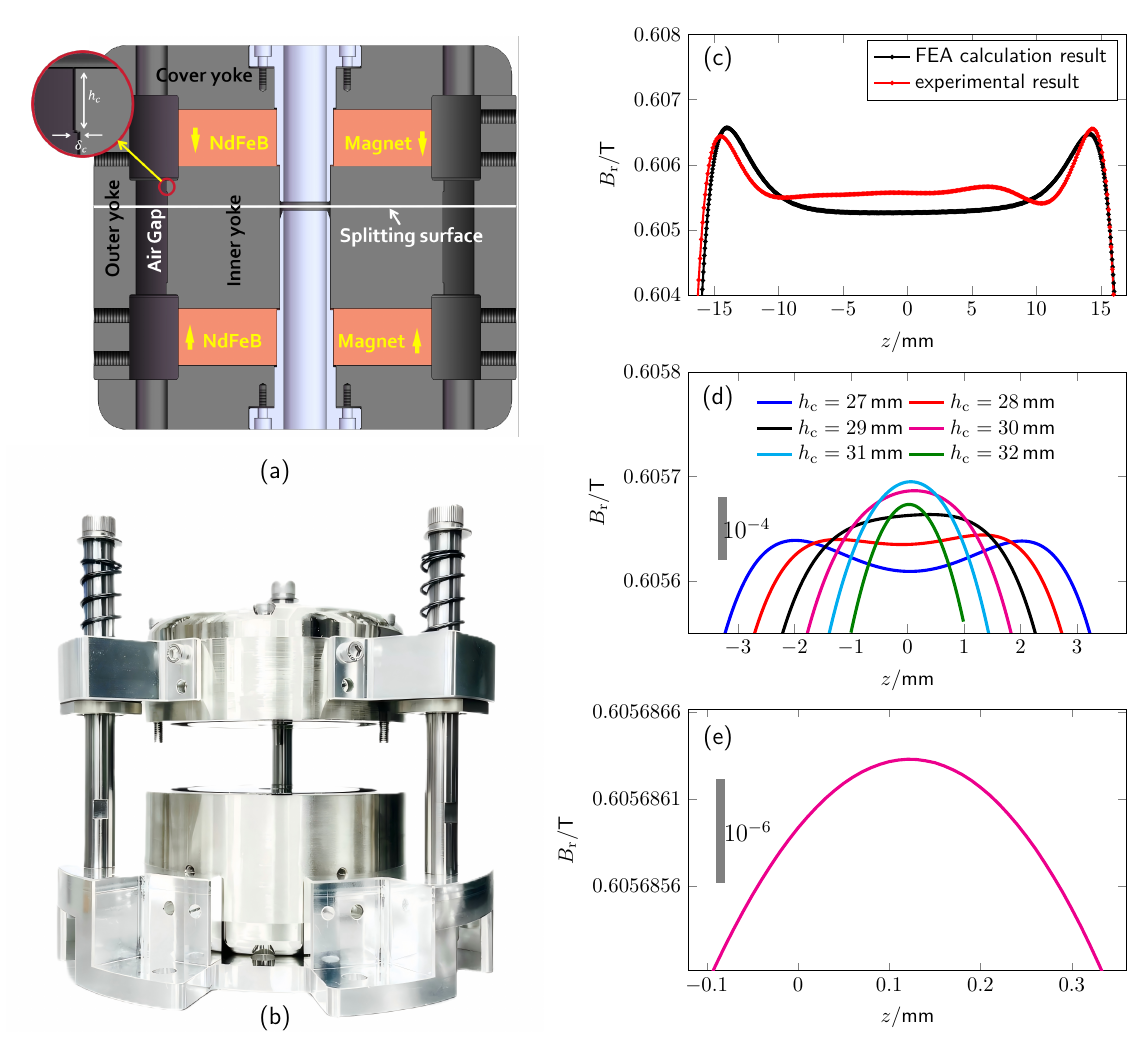}
\caption{Magnet-coil system design of KBmini. (a) presents the magnet's structure, with arrows indicating the magnetization direction. {The red circled area indicates a ring-shaped compensation structure on both ends of the magnet.} (b) depicts the system in an open, magnetically levitated state. (c) compares the measured magnetic flux density profile with results from a finite element analysis (FEA). (d) shows how the profile shape varies with different coil heights. (e) displays the magnetic profile within the velocity measurement range for a selected coil height of $h_c=30$\,mm.}
\label{magnet_ex}
\end{figure}

Similar to \cite{li2024magnet}, two additional new features are integrated into the magnet system: 
\begin{enumerate}
\item A ring compensation with a rectangular cross-section
(0.3\,mm radius increase, 4\,mm in height) is added to both ends of the inner yoke. Theoretical calculation shows that this proposed compensation can increase the uniform field range by over 30\%. 

\item An optimal splitting surface is designed at $z = 11$\,mm. The magnetic force of two {sections}, shown in figure \ref{magnet_ex}(b) is attractive ($\approx 92$\,N) and once it is opened by a few milimeters, the force becomes repulsive, which allows a complete magnetic levitation of the upper section with the help of a guide system. The levitation distance is above 40\,mm, allowing very convenient coil operations. 
\end{enumerate}

The key parameters of a typical Kibble balance magnetic circuit are the magnetic field strength within the air gap and the length of the uniform region of the field. These parameters are typically measured by using a gradient coil \cite{NISTshimming}. Figure \ref{magnet_ex}(c) illustrates the comparison between the measured and simulated magnetic field strengths within the air gap of the magnet. The red curve denotes the experimentally obtained data, whereas the black curve represents the results derived from FEA calculation. 
The measured magnetic field strength within the air gap is 605.5\,mT. The uniform magnetic field region, defined by the condition $\Delta B/B$ $<$ $1\times 10^{-3}$, has a measured length of 21.9\,mm (compared to 23.0\,mm in simulation). Further details regarding the magnet system and air gap magnetic field compensation can be found in \cite{li2025approachrestoringmagneticfield}.

In oscillation-{style} Kibble balances, the movement range of the flexure hinge is usually very limited, yielding a relatively low moving velocity. To achieve a considerable induced voltage $U$ for enhanced {velocity phase measurement accuracy}, a high $Bl$ value is preferred. {Given the limited moving range along the vertical, a taller coil that utilizes as much as the percentage of the uniform field range helps to produce larger $Bl$ and lower systematics, such as the thermal effect.} Considering the non-uniformity of the magnetic field within the air gap, it is essential to calculate the average magnetic field for coils of different heights at various positions within the air gap. This analysis helps to determine the optimal range of motion and the appropriate coil height.

As illustrated in figure \ref{magnet_ex}(d), the average magnetic profile is plotted for different coil heights \( h_c \) at various positions \( z \) within the air gap. Furthermore, figure \ref{magnet_ex}(e) presents the average magnetic profile for \( h_c = 30 \) mm over a 0.4\,mm range of motion, where the magnetic profile fluctuates a few parts in \( 10^{6} \), demonstrating excellent uniformity.

A major systematic effect related to the magnet-coil system is the current effect, written as
\begin{equation}
    (Bl)_w=(Bl)_v(1+\alpha I+\beta I^2),
    \label{current}
\end{equation}
where $(Bl)_w$, $(Bl)_v$ are geometrical factors obtained in the weighing and velocity measurements, $\alpha$ and $\beta$ are, respectively, the linear and nonlinear coefficients. 

\begin{figure}[tp!]
\centering
\includegraphics[width=0.65\textwidth]{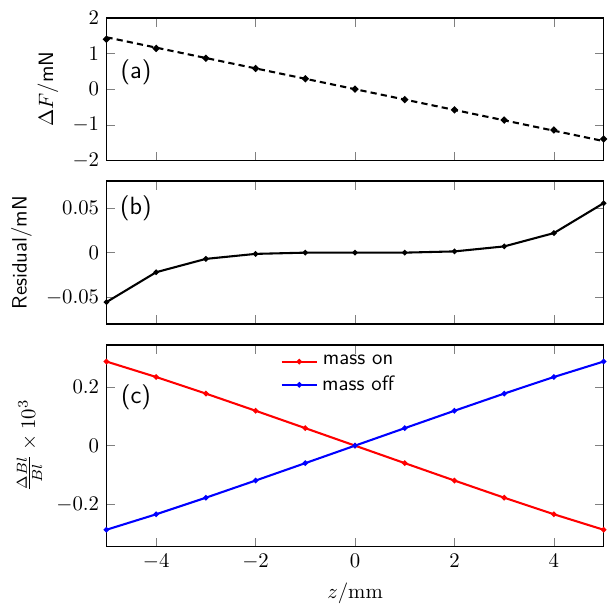}
\caption{{Current effect of the KBmini coil obtained by FEA calculations.} (a) displays the inductive force as a function of vertical position $z$, expressed as $\Delta F = \frac{I^2}{2} \frac{\partial L}{\partial z}$. The coil configuration corresponds to 24.5 ampere-turns, equivalent to the operating condition for a 1\,kg standard mass load with $I = 3.5$ mA and $N = 7000$ turns. {Although the FEA was performed in $\pm 5$\,mm, as the coil moves in the sub-millimeter range, the linear fit (the dashed line) was conducted in the range of $|z|\leq1$\,mm}. (b) shows the residual inductive force after subtracting a linear component fitted within the $\pm 1$ mm range. (c) presents the relative magnetic field variation along $z$ for both positive and negative coil currents $\pm I$. 
}
\label{currenteffert}
\end{figure}

The linear term, $\alpha I$, arises from the coil self-inductance force~\cite{CoilCurrent_Li2018,li2017permanent,NISTshimming}. In principle, this term can be eliminated through the use of symmetrical currents during mass-on and mass-off weighing measurements. However, a significant correction to this approach stems from coil displacement variations between mass-on and mass-off states. To evaluate this effect, we conducted an FEA simulation based on a 1 kg mass configuration, where permanent magnets were replaced with air in the model.

The resulting inductive force $\Delta F$ as a function of vertical position $z$ is shown in figure~\ref{currenteffert}(a). This force acts as a restoring force, tending to push or pull the current-carrying coil toward the magnet center. As illustrated in figure~\ref{currenteffert}(b), the $\Delta F(z)$ relationship exhibits good linearity within the central region ($|z| < 2$ mm). For the 1 kg load case with weighing currents of $\pm 3.5$ mA (generating a nominal force of 4.9 N), figure~\ref{currenteffert}(c) shows the relative magnetic field variation between mass-on and mass-off measurements. The absolute slope in the central (weighing) region is $5.9 \times 10^{-5}$/mm, corresponding to a relative correction of $5.9 \times 10^{-8}$ per micrometer of coil displacement change between mass-on and mass-off states.
 
As identified in \cite{Li_MagneticUncertainties,liu2025}, compared to other nonlinear magnetic errors~\cite{Nonlinear_2013,Nonlinear_2014,hysteresis2020}, thermal-magnetic effects present a significant source of uncertainty for two-mode tabletop Kibble balances. The underlying mechanism arises from ohmic heating in the coil, which occurs exclusively during the weighing phase. This creates a temperature differential between the weighing and velocity measurements. The resulting temperature fluctuation is coupled to the permanent magnet, which typically features a negative temperature coefficient (e.g., $T_c \approx -1 \times 10^{-3}$/K in the KBmini magnet system), leading to a corresponding change in the magnetic field.

\begin{figure}[tp!]
\centering
\includegraphics[width=0.65\textwidth]{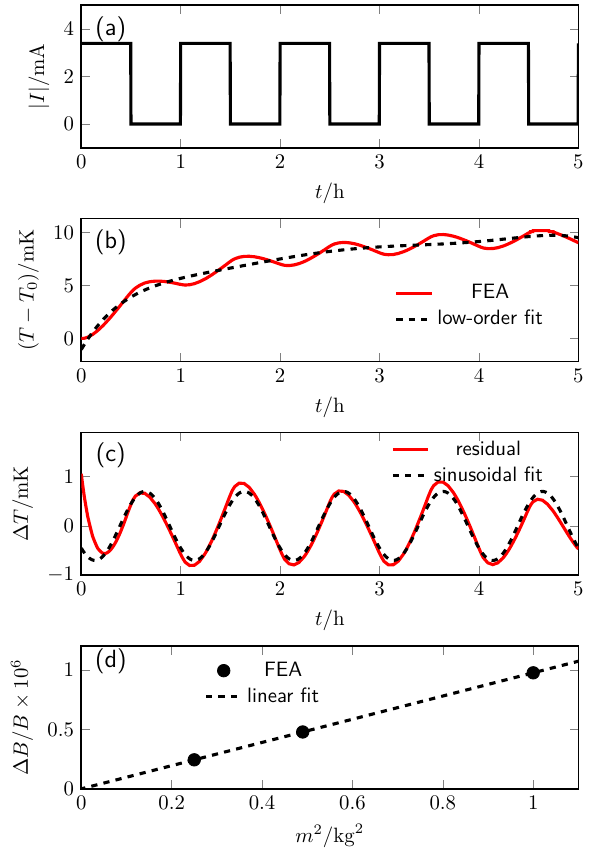}
\caption{Thermal effects in the KBmini magnet system. (a) Current input to the magnet coil versus time. (b) Temperature {($T$) of the ring magnet} rise relative to the initial temperature (\(T_0\)) as a function of time. The red solid curve and black dashed curve represent results from FEA simulation and a low-order polynomial fit model, respectively. (c) Residual temperature fluctuation (red solid), obtained by subtracting the low-order fit prediction from the FEA data. A sinusoidal fit to the residual is shown as a black dashed curve. (d) The thermal-magnetic effect exhibits a linear relationship with the square value of the test mass.}
\label{T}
\end{figure}

Based on the methodology outlined in \cite{liu2025}, an FEA was conducted to simulate heat transfer from coil heating. The input current profile applied to the coil is shown in figure~\ref{T}(a). Each weighing phase has a duration of 30 minutes with a current magnitude of $|I| = 3.5$\,mA, producing a magnetic force of $\pm 4.9$\,N. This results in a coil heating power of approximately 30\,mW. The subsequent velocity phase also lasts 30 minutes, during which the coil current is maintained at zero. 
Figure \ref{T}(b) presents the temperature {($T$) of the ring magnet} rise relative to the initial temperature $T_0=25\,^{\circ}\mathrm{C}$, where the red solid line represents the raw FEA results, and the black dashed line corresponds to a 6th-order polynomial model. The residual between these datasets is fitted with a sinusoidal function having the same period as the current profile, as shown in figure \ref{T}(c). This sinusoidal fit yields a temperature fluctuation amplitude of 0.7\,mK. By averaging the temperature {shown in figure \ref{T}(c)} over the weighing and velocity measurement intervals separately, a mean temperature difference of 0.97\,mK is obtained. These thermal variations consequently introduce a systematic uncertainty of $9.75 \times 10^{-7}$ under a 1\,kg load. {Note that in practice, the thermal change inside the permanent magnet is difficult to measure due to an unknown delay related to the magnet's thermal capacity, and hence corrections based on thermal measurement are not recommended. } 

Given the significance of this bias, methods for its suppression or elimination must be implemented. One viable approach is to experimentally determine the bias introduced by this effect and apply a corresponding correction. The thermal-magnetic effect manifests as a second-order term, specifically $\beta I^2$, in the current effect described by equation (\ref{current}). This implies that by conducting measurements with different known mass values, the effect can be extracted through a linear fit of the differential mass values $\Delta m$ (the difference between measured and known values) against $I^2$, or alternatively, via a quadratic fit of $\Delta m$ against $I$. To validate this, we performed additional FEA calculations for mass weighings of 700 g and 500 g. Figure~\ref{T}(d) shows the relative magnetic field change due to coil heating as a function of $m^2$, which exhibits a linear relationship passing through the origin.

\subsection{Coil motion control}

Kibble balances primarily employ two distinct motion strategies: moving-coil and moving-magnet configurations. In the moving-coil design, the magnet is stationary while the coil moves through the magnetic field. Conversely, the moving-magnet strategy uses a fixed coil, with the magnet's motion altering the field experienced by the coil~\cite{Li_2017, UME2023, THUdesign2022}. Flexure-hinge-based Kibble balances offer highly precise and compact weighing measurements. However, a key challenge is the hinge's limited motion range, typically under 1\,mm, which necessitates rapid movement and data acquisition when used directly for coil movement. This constraint also complicates motion control, as traditional position/velocity-tracking loops must minimize acceleration and deceleration displacements to reach the target state quickly.

Recent developments in flexure-based tabletop Kibble balances at NIST~\cite{NISTTabletop2024} and PTB~\cite{PTB2023} have demonstrated that 1) flexures can endure motion ranges exceeding 1\,mm, and 2) sinusoidal excitation can achieve sufficiently high coil velocities on the order of sub-millimeter per second. These findings are significant, as they eliminate the need to move either the weighing unit~\cite{METAS2022,LNE2025,KRISS2020} or the magnet~\cite{NIM2023,UME2023,THUKB2023}. We note that a sinusoidal motion profile produces a continuously varying velocity and thus an alternating induced voltage. The high-precision measurement of such AC signals requires a quantum-based reference voltage from a Programmable Josephson Voltage Standard (PJVS)~\cite{PJVS} or a Josephson Arbitrary Waveform Synthesizer (JAWS)~\cite{JAWS} system. Furthermore, frequency-dependent effects in AC measurements may introduce non-negligible corrections~\cite{PTB2025}.

The KBmini system presented in this work implements a {multi-harmonic excitation} approach to motion generation. This technique injects harmonic currents into the drive coil, effectively reshaping the conventional sinusoidal trajectory into a square-wave-like profile. {By leveraging the intrinsic dynamics of the flexure hinge, the method achieves near-uniform velocity over a full movement cycle.}

The underlying principle can be explained as follows. {The flexure hinge is modeled as a single-frequency oscillator characterized by a linear restoring force. } Its free oscillation is governed by the differential equation: 
\begin{equation}
\frac{\mathrm{d}^2 z}{\mathrm{d} t^2} + 2\varsigma\omega_0 \frac{\mathrm{d} z}{\mathrm{d} t} + \omega_0^2 z = 0,
\end{equation}
where $\omega_0$ is the natural frequency and $\varsigma$ the damping ratio.
When a sinusoidal drive current $I_0 \sin(\omega t)$ is applied to the coil of a linear actuator within a small oscillation range, the equation of motion becomes:
\begin{equation}
\frac{\mathrm{d}^2 z}{\mathrm{d} t^2} + 2\varsigma\omega_0 \frac{\mathrm{d} z}{\mathrm{d} t} + \omega_0^2 z = \rho(Bl)_A I_0 \sin(\omega t),
\label{eq:nonlinearOS}
\end{equation}
where $(Bl)_A$ represents the actuator's geometric factor and $\rho$ is a normalized ratio incorporating the system inertia and force implementation beam ratio.

The steady-state solution to equation (\ref{eq:nonlinearOS}) consists of sinusoidal displacement and velocity responses with amplitudes $z_0$, $v_0$ and phase shifts $\varphi$, $\varphi'$:
\begin{equation}
z(t) = z_0 \sin(\omega t + \varphi), \quad v(t) = v_0 \cos(\omega t + \varphi').
\label{eq:OSolution}
\end{equation}

{The target is to achieve an ideal square wave for the velocity $v(t)$. Since $v(t)$ is defined as a cosine function with simple harmonic oscillation, as presented in equation (\ref{eq:OSolution}), the ideal square $v(t)$ waveform is symmetrical to the vertical axis and} can be decomposed into a fundamental frequency component ($\omega$) and a series of even-order harmonics ($2\omega$, $4\omega$, $\dots$). According to equation (\ref{eq:OSolution}), generating a square-wave-like velocity profile requires the introduction of even-order harmonics into the velocity response. This is achieved by injecting odd harmonics ($3\omega$, $5\omega$, $\dots$) into the driving current. By appropriately adjusting the amplitudes and phases of these harmonics, the velocity waveform can be shaped to exhibit extended flat regions in the time domain, thereby closely approximating a constant velocity profile.

To validate this concept, we conducted an experiment in which a current was applied to the coil of the magnet within the weighing unit, and the resulting displacement was measured using the capacitive sensor from the weighing feedback loop. It should be noted that due to the asymmetric arm lengths on either side of the hinge, the actual velocity of the main coil is approximately one-fifth of the measured displacement-derived value.

In the experiment, current signals with frequencies of 1.0\,Hz, 3.0\,Hz, 5.0\,Hz, 7.0\,Hz, and 9.0\,Hz were applied sequentially. Through iterative optimization, the current amplitudes ratio was determined to be $A_1:A_3:A_5:A_7:A_9=1:1:0.6:0.36:0.14$, with corresponding phase values of 0\,rad, -0.8\,rad, -2.3\,rad, -3.8\,rad, and -5.3\,rad. {It should be noted that the configuration presented above represents merely one viable set of optimal parameters. Due to the inherent nonlinearity of oscillatory systems, no universal analytical model has been found to precisely determine the requisite amplitudes and phases for harmonic components. A practical methodology involves the stepwise introduction of odd harmonics, beginning with the lower frequencies (e.g., $3\omega$, $5\omega$, ...) and progressively adding higher-order components, with the final parameters typically refined through iterative adjustment.} Subplots (a) through (e) in figure \ref{non_liner} demonstrate the progressive improvement in velocity flatness achieved through the sequential introduction of additional harmonics with the shown example. During the combining of these components, the maximum current is limited to $\pm$0.6\,mA. 

\begin{figure*}[tp!]
\centering
\includegraphics[width=0.85\textwidth]{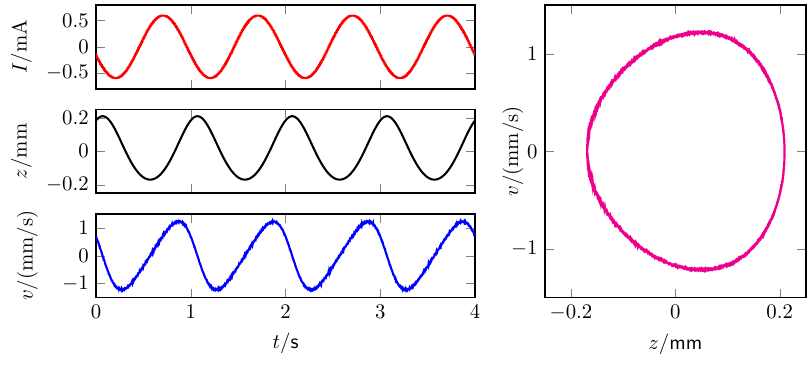}\\
(a)\\~\\
\includegraphics[width=0.85\textwidth]{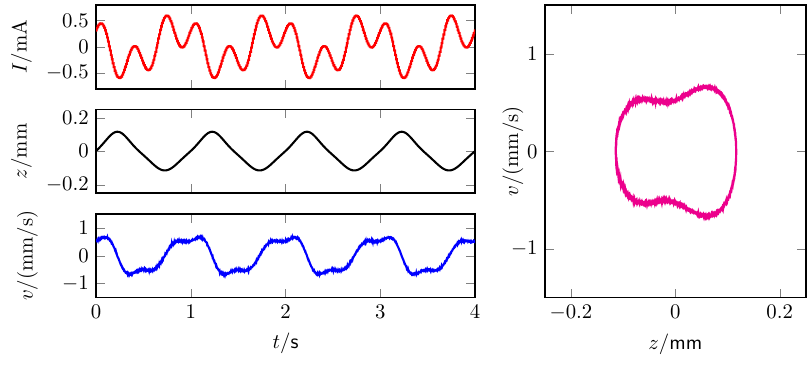}\\
(b)\\~\\
\includegraphics[width=0.85\textwidth]{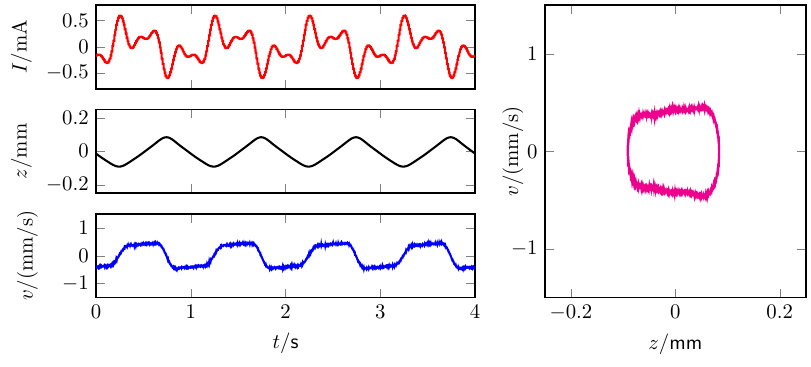}\\
(c)\\
\end{figure*}

\begin{figure*}[tp!]
\centering
\includegraphics[width=0.85\textwidth]{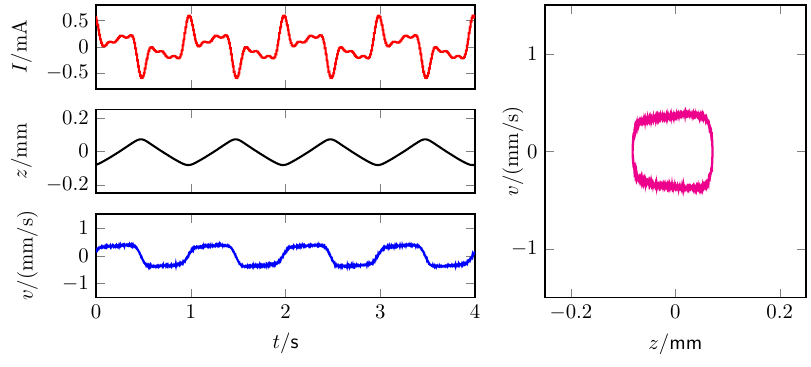}\\
(d)\\~\\
\includegraphics[width=0.85\textwidth]{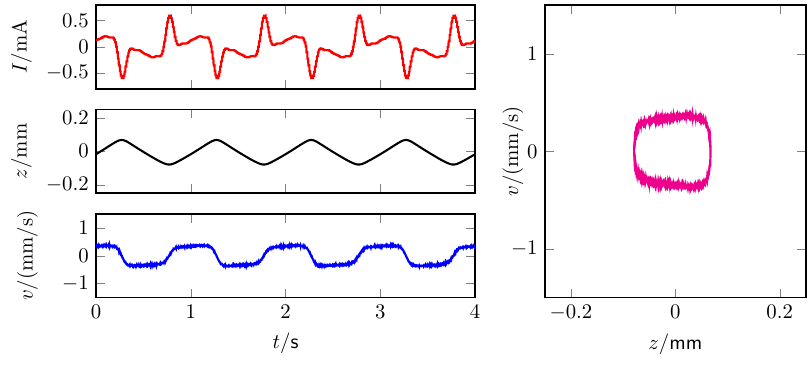}\\
(e)\\~\\
\includegraphics[width=0.85\textwidth]{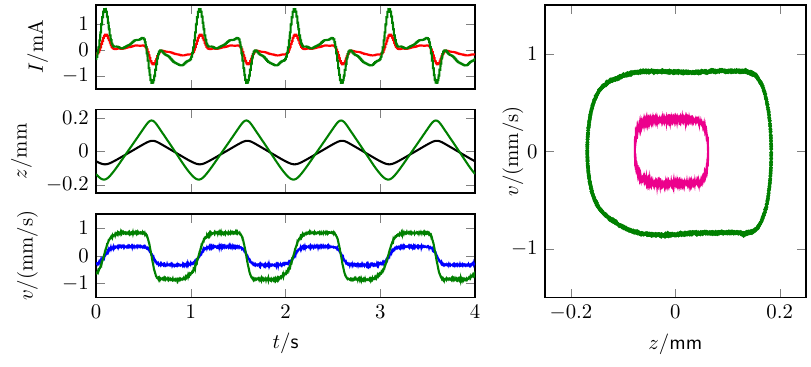}\\
(f)\\
\caption{An experimental demonstration of the {multi-harmonic excitation}. Subplots (a) to (f) respectively show the oscillations with different harmonics: (a) $\omega$; (b) $\omega$, $3\omega$; (c) $\omega$, $3\omega$, $5\omega$; (d) $\omega$, $3\omega$, $5\omega$, $7\omega$; (e) $\omega$, $3\omega$, $5\omega$, $7\omega$, $9\omega$; (f) $\omega$, $3\omega$, $5\omega$, $7\omega$, $9\omega$, $2\omega$.  }
\label{non_liner}
\end{figure*}

The standard deviation of velocity, computed over a 0.3\,s window centred at the midpoint of each half-cycle, was reduced by an order of magnitude—from 0.04 mm/s in the initial sinusoidal state to 0.004 mm/s.
However, an undesired waveform tilt was observed, as illustrated in figure \ref{non_liner}(e). Analysis revealed that this tilt stems from the quadratic spatial nonlinearity of the flexure's restoring force and the nonlinearity of the magnetic circuit's $(Bl)_A$ factor. To compensate for this effect, a second harmonic (2.0 Hz) was introduced into the correction system, with an optimized amplitude of 0.026\,mA and a phase matching that of the fundamental wave. As shown in figure \ref{non_liner}(f), this adjustment resulted in a velocity profile that maintained high uniformity over more than 60\% of the motion cycle ($\Delta v / v < 5\%$).
It can be seen that under combined excitation of the 1.0\,Hz fundamental wave and the harmonic components (2nd, 3rd, 5th, 7th and 9th), the system achieved a flat velocity. The stable velocity $\approx0.3$\,mm/s is much lower compared to the peak velocity with only the fundamental component. This can be fixed by simply amplifying the drive current, and the green curves in figure \ref{non_liner}(f) show a case with an average velocity of 0.8\,mm/s.

\section{Conclusion}
\label{sec04}

This paper has presented the design of KBmini, a tabletop Kibble balance targeting E2-class mass calibration {from 1\,g to 1\,kg}. The system incorporates several ideas for further minimizing the tabletop Kibble balance designs. The integrated weighing unit with capacitive sensing and the comprehensive coil positioning system provide the necessary precision for high-accuracy measurements. The compact BIPM-type magnet system, with an outer diameter of 150\,mm and total weight of 14\,kg, achieves a high magnetic flux density of approximately 0.6\,T in the air gap while featuring an optimal splitting surface that enables convenient coil access through magnetic levitation. For a two-mode measurement scheme, the thermal-magnetic effect is shown to be considerable with high load, and experimental corrections in this case must be applied {as shown in figure \ref{T}}. 

The most significant contribution of this work is the development of a {multi-harmonic excitation} approach for velocity phase measurements. By strategically injecting odd-order harmonics into the driving current, we have demonstrated the transformation of sinusoidal coil motion into a square-wave-like velocity profile with significantly improved flatness. Furthermore, the introduction of a second harmonic component effectively compensates for waveform tilt caused by mechanical and magnetic nonlinearities, enabling uniform velocity over more than 60\% of the motion cycle with velocity variations of only a few percentages.
The {multi-harmonic excitation} strategy leverages the intrinsic dynamics of the hinge system while requiring only minimal corrective currents, making it particularly suitable for compact implementations. This approach maintains the advantages of flexure-based systems—high precision and compactness—while overcoming traditional limitations in velocity control and signal measurement.

Future work will focus on the complete integration of all subsystems and comprehensive uncertainty evaluation toward the target E2-class accuracy. The successful implementation of KBmini will make quantum-based mass realization accessible to a wider range of calibration laboratories and industrial applications, supporting the dissemination of the revised SI in practical mass metrology.

\section*{Acknowledgment}

Shisong Li would like to acknowledge the financial support from the National Natural Science Foundation of China (No. 52377011). The authors would like to thank Kibble balance colleagues from NIST and BIPM for valuable discussions. 

\section*{References}

\end{flushleft}
\end{document}